# A million cycles in a day: enabling high-throughput computing of lithium-ion battery degradation with physics-based models

Ruihe Li[a,b †], Simon O'Kane[a,b], Jianbo Huang[c], Monica Marinescu[a,b], Gregory J Offer[a,b ‡]

[a] Department of Mechanical Engineering, Imperial College London, UK

[b] The Faraday Institution, UK

[c] Cloud Valley, NO.1008 Dengcai Street Xihu District, Hangzhou, China

[†] r.li20@imperial.ac.uk, [‡] gregory.offer@imperial.ac.uk

## Abstract

High-throughput computing (HTC) is a pivotal asset in many scientific fields, such as biology, material science and machine learning. Applying HTC to the complex physics-based degradation models of lithium-ion batteries enables efficient parameter identification and sensitivity analysis, which further leads to optimal battery design and operating conditions. However, running physics-based degradation models comes with pitfalls, as solvers can crash or get stuck in infinite loops due to numerical errors. Also, how to pipeline HTC for degradation models has seldom been discussed. To fill these gaps, we have created ParaSweeper, a Python script tailored for HTC, designed to streamline parameter sweeping by running as many ageing simulations as computational resources allow, each with different parameters. We have demonstrated the capability of ParaSweeper based on the open-source platform PyBaMM, and the approach can also apply to other numerical models which solve partial differential equations. ParaSweeper not only manages common solver errors, but also integrates various methods to accelerate the simulation. Using a high-performance computing platform, ParaSweeper can run millions of charge/discharge cycles within one day. ParaSweeper stands to benefit both academic researchers, through expedited model exploration, and industry professionals, by enabling rapid lifetime design, ultimately contributing to the prolonged lifetime of batteries.

## Introduction and Motivation

Lithium-ion batteries (LIBs), frequently used in electric vehicles, smartphones, and renewable energy systems, are arguably the most important energy storage devices in the modern world. Despite their advancement compared with their counterparts, LIBs still face challenges such as worsening performance in the long-term due to various degradation mechanisms. One powerful tool to understand these complex degradation mechanisms is physics-based models, which have been extensively developed in recent years.

Physics-based models can be categorised into beginning of life (BOL) models and degradation models. The BOL model normally refers to the classical pseudo two-dimensional model (P2D), also known as the Doyle-Fuller-Newman model (DFN) [1]. The P2D model works by solving partial differential equations (PDEs) in a one-dimensional space containing two electrodes and one separator. At each point of the electrode regions, the solid-phase diffusion equation is solved in a second radial dimension. The degradation models describe the degradation mechanisms of LIBs with partial differential equations based on a deep understanding of the physical processes. Degradation mechanisms that have been modelled so far include solid-electrolyte interphase (SEI) layer growth [2], crack growth [3], loss of active material (LAM) [4], lithium plating [5], and solvent consumption [6], which have been systematically reviewed by Reniers *et al.* [4] and Edge *et al.* [7].

High-throughput computing (HTC), i.e., running massive numbers of cases of high-fidelity numerical models, is a common yet important research method in many fields. For example, in the field of biology, toolkits like GROMACS [8] have been proposed to enable the high-throughput molecular dynamics simulation of biomolecules. In material science, high-throughput computing is widely used in screening candidate structures with targeted performance [9-11]. However, there has not been any kind of such tool tailored for the physics-based degradation models for LIBs. The importance of HTC for is two-fold. Firstly, HTC is an indispensable way to parameterise the physics-based degradation models. Unlike the parameters of physics-based BOL models, which have been measured through carefully designed individual experiments, parameters of degradation models are normally difficult or almost impossible to measure directly [3, 12, 13]. The only way to identify these parameters is to fit the modelling results to experimental data, which requires using HTC to sweep over the possible ranges and finding the result with the least error. Secondly, the massive numbers of input-output pairs produced by HTC can be used to train surrogate models via machine learning algorithms. Alipour *et al.* [14] trained a surrogate model using ~200 cases of synthetic data from the DFN model coupled with a thermal model and four degradation sub-models. With the trained surrogate model, 56 model parameters were identified and their sensitivity was analysed. Appiah *et al.* generated 30,000 combinations for 15 input parameters and obtained 23,074 outputs using a DFN model with an SEI model. The input-output pairs were then fed into a Gaussian process regression model to facilitate global and local sensitivity analysis and to explore the relationship between SEI thickness and irreversible charge loss.

However, using HTC (or parameter sweeping) to parameterise physics-based degradation models is rarely mentioned in literature. Most modelling papers just give the best fit found directly as a *fait accompli*. Details on how to conduct or pipeline such a large amount of simulation cases are missing. The fact that not all input cases converge in Ref. [15] reveals one important pitfall in running high-fidelity numerical models, i.e., the solver may crash due to numerical errors. To the best of our knowledge, we found only one article [16] by the creators of the P2D model discussing two common problems encountered when using it and how to resolve them. We have noticed lots of questions raised in the PyBaMM GitHub repositories [17] which are related to these pitfalls.

Therefore, the battery modelling community urgently needs guidelines on how to automate dealing with simulation problems in order to fully utilize HTC for the complex physics-based degradation models. Such guidelines would be as valuable to modellers as those on how to make coin cells and single layer pouch cells [18] are to experimentalists. To fill this gap, we first explain the common pitfalls, including examples of solver crashes and numerical challenges that can occur during physics-based modelling (with any software) and provide possible solutions in the method section. We then introduce ParaSweeper, a Python script that can run many simulations at once, identify those that crash and automatically run the best possible alternatives. Finally, we show how this script can efficiently run 1000 cases and manage the output for sensitivity analysis and model fitting using the high-performance computing platform of Imperial College London.

**Pitfalls in physics-based modelling**

Physics-based models involve solving differential equations, introducing the potential for numerical problems and model crashes . To ensure that the model faithfully replicates the real-world scenarios, specific limits should be imposed on some variables, such as voltage for the case of LIBs. This is accomplished by setting events and cut-offs. In the following, we will elaborate common events, cut-offs, and numerical errors based on an open-source platform PyBaMM [19]. Compared with its counterparts [20], PyBaMM implements physics-based models in a modular fashion, enabling easy implementation of new models and numerical methods. A large variety of novel physics has been added to PyBaMM [5, 6, 21-23] and it now has a fast-growing community. Similar events and cut-offs will be set by other battery modelling platforms, and the numerical problems and solutions discussed here apply to other physics-based models solving PDEs.

**Events and cut-offs**

Two main purposes of events and cut-offs are: (1) switching the simulation to the next step, for example, from constant current to constant voltage; (2) noticing unphysical modelling results and terminating the simulation. Theoretically, if the physics-based model can genuinely reproduce all the behaviour of the true object, all the internal states and parameters are always physically sensible. However, some of the relevant physical processes, such as SEI formation, are still not fully understood. Also, some physical processes are less important under certain circumstances and therefore are omitted for the sake of computational efficiency. All physics-based models have fundamental flaws and may give unphysical results.

Common events for a virtual LIB include concentration cut-offs, voltage cut-offs, porosity cut-offs, current cut-offs, and temperature cut-offs. Sometimes these cut-offs just lead to a normal switch to the next step in an experiment. However, sometimes these can lead to early termination of the modelling.

Voltage cut-off is the most straightforward one, referring to the fact that the voltage of the cell cannot exceed a certain range assuming proper protection is in place. Current cut-off is normally used during constant voltage hold, and the model will be switched to the next step when a certain current is reached.

Concentration cut-off for an electrode particle means that its concentration cannot exceed the maximum/minimum values that can intercalate/de-intercalate from the electrode materials. In reality, if the particle surface concentration exceeds the limit, the exchange current density should become zero, increasing the reaction overpotential and triggering the voltage cut-off. However, adding this physical constraint to the exchange current density may sometimes be difficult, which makes concentration cut-off necessary.

The concentration cut-off limits for the electrolyte are zero cut-off and saturation cut-off. Concentration here refers to the concentration of salt, or of Li$^+$. It is unphysical to have negative species concentration or concentrations above the saturation concentration in the electrolyte. A zero-salt concentration cut-off therefore makes physical sense. However, in some cases, allowing the salt concentration to go negative (by a relatively small amount and for a short time) can prevent numerous simulation terminations and therefore should be allowed if the error introduced is small. Considering that no model is perfect and that strictly constraining the concentration make simulation difficult, this is a trade-off. A saturation limit, beyond which the salt will precipitate, also makes physical sense. However, adding the precipitation and dissolution dynamics into a LIB model will increase computational resources. As precipitation is expected to occur only at extreme cases, it is not usually implemented in the model.

The input parameters of DFN models are measured through carefully designed experiments on a series of samples of concentrations at different temperatures, then fit to an empirical form of mathematical functions. Therefore, it is important to remember that outside the measured ranges, these functions may give unphysical values. In one of our example notebooks, where the negative porosity is assumed to be reduced from 0.25 to 0.1 due to SEI layer growth, we showed that upon 1.8C discharge, using a polynomial form of electrolyte properties will give extremely high value of diffusivity and conductivity at a salt concentration of 5M, at which salt precipitation should occur. Such polynomial extrapolation can give a discharge capacity twice as large compared to assuming a flat linear extrapolation. Unfortunately, the saturation limit for electrolyte has not been implemented in the mainstream LIB simulation tools. Using a flat linear extrapolation is probably a safer assumption than a polynomial as a temporary solution until more work has been done at high concentrations [24].

Porosity cut-off is another common event in physics-based modelling of LIBs. It can be induced by extremely high external pressure and high temperature. However, the more common scenario is pore-clogging in the negative electrode due to side reaction deposits such as SEI growth and lithium plating during long-term usage. For the P2D model which has only one dimension in the thickness direction, zero porosity will completely shut down the current and make the simulated cell die. In a real cell, zero porosity in one region will cause abnormally high current density in the surrounding regions, causing those regions to degrade faster and possibly reach zero porosity as well.

**Numerical problems, solver errors**

Physics-based models' reliance on PDEs means that numerical problems may occur if the PDEs are mathematically unstable. Numerical problems include warnings and solver errors. In PyBaMM, warnings appear when an event has been hit and the PyBaMM is redoing the calculation to find out exactly when the event happens. If there are a lot of warnings, it means the model is approaching the limit of what it can do and must work extremely hard. But the result may still be correct. Solver errors may occur either due to the limitations of the solver itself or if the model is mathematically unstable. If they occur at beginning of life (occur at t=0), it is often necessary to change (slightly or massively) the parameter values to make it run.

One special type of issue in PyBaMM is "Experiment is infeasible." If it happens at t=0 or on the first cycle, it usually means that the experiment cannot be run within the cell's safety limits. If it happens in later cycles, the solution up until this point is returned, and the reason is usually that the cell died due to degradation. It could be completely dead or still work for lower current.

## Methods to resolve pitfalls and accelerate simulation

The authors have accumulated lots of precious experiences after years of debugging physics-based models, which serve as methods to resolve the pitfalls mentioned above and develop ParaSweeper.

In terms of resolving the problems mentioned above, the solutions will be:

(1) First, check whether the BOL parameters fall within the physically plausible range. Reliable sources of those data include Ref. [25-27] and platforms like LiionDB [27, 28] and Voltt [29]. This is much more difficult for degradation parameters, but checking the parameter ranges in the corresponding papers with that degradation mechanism can provide a good starting point. It is common for the virtual cell to fail or die quickly due to high degradation parameter values.

(2) Double check the balancing of the cell and ensure the voltage is correct at BOL.

(3) For voltage cut-off, it may be due to impedance being too high. The most common causes of high impedance are low porosity and low electrolyte conductivity, but the impedance is affected by many factors and more investigation will be needed. It could also be that the cell is already at 100% SOC, in which case if it is charged, warnings or errors can occur due to solid concentration limits being exceeded. PyBaMM has been designed such that when using the Experiment class and the protocol is infeasible, it will hit the voltage limit immediately, then this step will be skipped and return an empty solution. This is extremely helpful in GITT characterization for aged cells as the impedance grows a lot.

(4) The problem of concentration cut-off (solid or liquid) may indicate that the concentration gradient is too high due to limited diffusivity or high current. Try to increase diffusivities or reduce the current.

(5) It is important to remember that the model is solved one time step after another. Some solvers will set a relatively long-time step to ensure short computation time (*dt_max* in CasADi solver). But if the internal states change too quickly, the solver may not be able to capture the changes and the solver does not converge. Many solvers will reduce the time step if they do not converge, but they also have a limit (*max_step_decrease_count* in CasADi solver) of how many times it makes the reduction before giving up and returning an error. Also, the standard of convergence can be changed slightly to make the model work sometimes (*rtol* in CasADi solver).

(6) There is also a similar trade-off in terms of space. In one example notebook, we show that if non-linear solid diffusivity is used, the number of mesh points along the radius direction must be increased to capture the concentration changes in the particle. Otherwise, the concentration distribution and therefore the voltage will be wrong. Modellers need to decide the optimal timestep and mesh points that can make sure the important aspects of the model are captured but the computational time is acceptable.

(7) For a solver error at t=0, try putting more mesh points in the region where the problem is stiff at t=0. For example, for particle cracking, put more mesh points at the particle surface where the crack originates. Inconsistent initial conditions may also occur if the electrolyte potential is manually given (e.g., COMSOL). But in PyBaMM, the electrolyte potential is calculated automatically from the open-circuit voltages.

(8) If you cannot reproduce the published results, check your notebook/script is exactly the same as the authors. Also check whether your software (and dependencies) installations are up to date.

(9) Very rarely, the solver can get stuck and run into an infinite loop, which will never return a solution. The only strategy to prevent this is a timeout function to stop the calculation after a pre-set time.

In the following are four methods to accelerate the simulation:

(1) Always use the minimum mesh points and maximum timestep possible. Change these based on the complexity of the problem. For example, we found that fewer mesh points in the particle are needed when particle cracking is disabled, or linear solid diffusivity is used.

(2) Save the minimum output variables needed and save them less frequently. We found that in many cases, storing the output data, instead of solving the PDEs, uses most of the computational resources. Think carefully about what is important and save only those results. This is especially important in degradation modelling when thousands of cycles are run. Some modelling tools allow the user to specify what output variables are saved. Regarding saving less frequently, PyBaMM has the *save_at_cycles* option, which only saves specified cycles over the total cycles (e.g., first, and last of the 1000 cycles). By saving fewer cycles, for example every 100 cycles out of the 1000 cycles (see *save_at_cycles* option in the example notebook), the memory requirement is reduced.

(3) However, if you are running 1000 cases of 1000 cycles and save every 100 cycles, you still have 10 cycles for each case. A potential way to further reduce the required memory is to delete solution object whenever possible and use custom dictionary. This can be combined with initializing the model from a previous solution. An example for PyBaMM is:

- Run several cycles, get the solution object,
- Extract necessary output variables,
- Use the solution to initialize a model,
- Delete the solution object and run the next round.

By doing so, only memory for one or two cycles of the solution is needed. This strategy will work in languages which allow users to allocate memory, such as C, but may not always work in Python which allocates memory automatically.

(4) Some programs allow the user to control the output frequency of results within one cycle or cycle step. In PyBaMM, this is done by setting a period for each step when specifying the experiment. However, if the period is too long, the output result will be coarse and fail to capture critical information. Therefore, the period must be chosen differently for each application.

If the solver gets stuck in an infinite loop, we may lose the results that were obtained before the solver got stuck. Therefore, it is advised to run the ageing protocol piece by piece. For example, if the total ageing protocol contains 1000 cycles, with reference performance test (RPT) every 100 cycles, try to run 100 cycles first, do the post-processing, then initialize the later modelling.

## Description of ParaSweeper

With all the above knowledge and experiences in mind, we have created the following script, ParaSweeper, to enable HTC for physics-based degradation modelling of LIBs. This script can be divided into 3 parts, as presented in Fig. 1. The model used in this work is the same as that used by O'Kane et al. [5], except the equations describing solvent consumption and electrolyte dry-out, which are taken from Li et al. [6]. Parasweeper is also compatible with other degradation models.

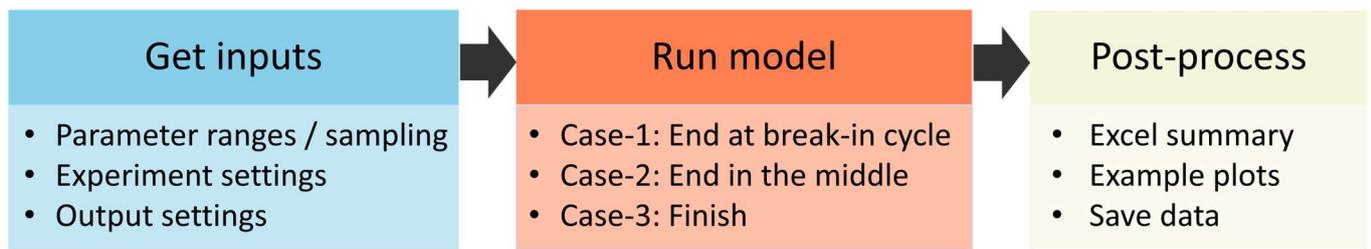

Fig. 1 Structure of the script

**Get inputs**

To start with, we need to determine what parameters we want to sweep over. For the BOL parameters, there are already quite lots of well-established methods to measure them and therefore they will only be within a narrow range. However, for degradation parameters, the range is much larger.

It is necessary at this point to clear up a common misconception about physics-based models; unlike equivalent circuit models, the parameters are fundamental properties of the cell that do not depend on current. The solid-phase diffusivity has been shown to vary with mechanical stress [30], but this is not the same thing. There is therefore only one set of parameters for each cell, and it is not physically realistic to use different parameter sets for different experiments run on the same cell, unless it can be shown that the cell has degraded in between the two experiments [31]. A possible use case for ParaSweeper would be to determine how much the degradation parameters very between different cells, but this would require degradation data for many different cells.

In the first part, we get all the necessary inputs for the simulations: (1) All parameters that determine the virtual cell, including BOL parameters, degradation parameters, and mesh settings. (2) Cycling conditions, i.e., what load we want to apply on the defined virtual cell (see *Cycling condition* in Supplementary Information). They can be cycling protocols like "charge at 1C until 4.2V", or current/load profile changing with time, or calendar ageing. (3) Output settings, to specify what variables we want to save during the later post-processing.

The main things to sweep are the model parameters and simulation settings. To be clear, we define all the inputs in one dictionary. As presented in *Example input dictionary* in Supplementary Information, two parts of inputs are specified. The first part is the model settings that cannot directly be given to the model, rather, this is the custom settings implemented in this script. This includes the parameter set (such as OKane2022) and mesh settings. For ageing simulations, it is also necessary to specify the total number of ageing cycles to run and the number of ageing cycles between two reference performance tests. The second part is the model parameters that will override the default values in a parameter set (in this case, OKane2022). Three parameters are specified in *Example input dictionary* in Supplementary Information, including two changing parameters and one unchanged parameter.

There are several ways to specify values for one single parameter (Fig. 2 (a)), i.e., manual picking, random sampling, linear spacing or log spacing, etc. The latter three methods require the upper and lower bounds. If the modeller is interested in more than one parameter, samples can still be generated for one parameter first, which are later combined together. Two ways have been implemented in ParaSweeper to do this, i.e., reference sampling and Latin hypercube sampling (Fig. 2 (b). Alternatively, we can randomly pick values of one specific variable for each combination (see the "all random" method in Fig. 2 (b)). Note that Latin hypercube can only be used when the sampling variables have the same number of elements. For two variables with different numbers of elements (e.g., one has 9 elements and the other has 3), reference sampling can be used.

**(a)  Sampling for one parameter**    **(b)  Combining multiple parameters**

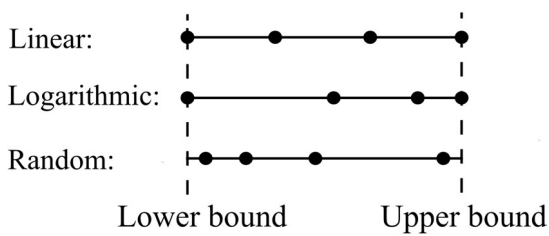 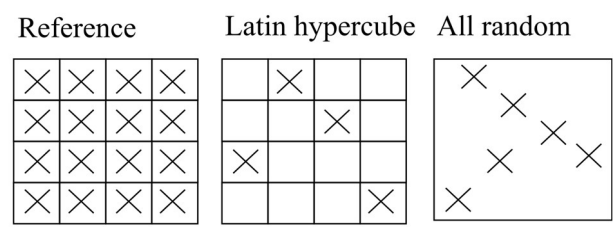

Fig. 2 Sampling methods for (a) one parameter and (b) multiple parameters.

After the sampling process, we will get multiple combinations (dictionaries in Python) with the same parameters (as keys in Python) but different values. The total number of combinations depends on the sampling method used. For example, if we have two parameters and have picked 4 values for each parameter (Fig. 2 (b)), the total combinations will be 16 and 4 for the reference and Latin hypercube sampling, respectively. For "All random", the number of total combinations can be random as well.

To record the input files and ensure we have submitted the desired values for simulation, we save these dictionaries into an CSV file. The CSV file will be loaded when running the model in the HPC system. Each row in the CSV file represents one case to run. In the last part of the Get Inputs Section, we define the output variables to be accessed during post-processing. These variables are normally the ones we care most about and intend to plot, such as terminal voltage, loss of capacity due to SEI, etc. The readers can refer to the script and the example output folder for more information.

**Run models**

The workflow of running the model is presented in Fig. 3. In Fig. 3, an example of running 1000 cases is presented. In Sweep_Age.csv, there are 1000 lines, corresponding to 1000 cases to run. Assuming each case requires 2 CPU and 15 GB of RAM and 8 hours, 1000 cases will require 2000 CPU and 15000 GB RAM for 8 hours. If all 1000 cases are submitted at once, it may threaten the normal operation of the HPC system. Therefore, it is advised to submit them in an array job with a small resource requirement in one single sub-job. Each sub-job will just read several lines from the generated CSV files. However, to avoid the input file Sweep_Age.csv being accessed simultaneously by two or more sub-jobs, which will produce problems, we have split the Age_1.csv into multiple files, naming from Bundle_1.csv to Bundle_250.csv. Within each Bundle_i.csv file, 4 lines are copied from Sweep_Age.csv. Each sub-job will just read their corresponding input file and run the model. These 250 files will then queue in the HPC job system and start whenever there are available resources.

To further accelerate the parameter sweeping process, the 4 cases within one sub-job in the above example are run in parallel (achieved by the multiprocessing package). Within one case which contains alternating N ageing cycles and one RPT cycle, we can set as follows so that only the results of first and the last cycles are saved:

*Sol_new = Simnew.solve(calc_esoh=False, save_at_cycles = N)*

The object *Sol_new* by default contains all the output variables that PyBaMM can calculate. However, we normally do not need that many variables and the solution object take quite a lot of memory. To save memory, we can specify desirable outputs and put it in a dictionary, so that the solution object can be deleted or overwritten. Meanwhile, we can use the *set_initial_conditions_from()* function so that the next simulation inherits the final states of the previous simulation.

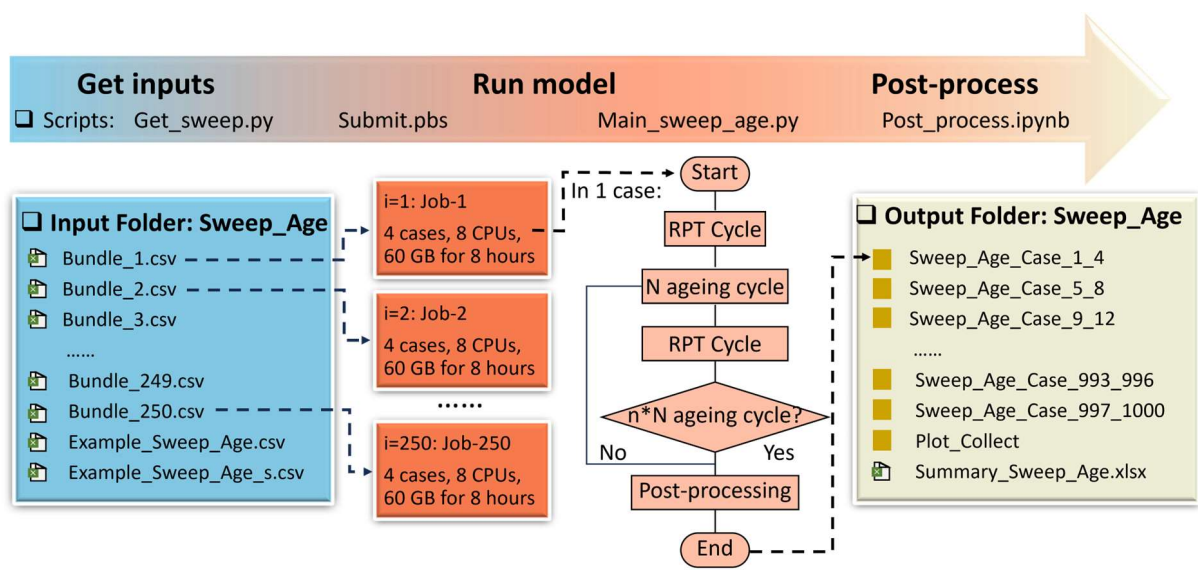

Fig. 3 Workflow of submitting jobs to HPC.

The number of cases to run in one sub-job depends on the capability of the HPC system used. Most HPC systems have guidance explaining what jobs the system can handle, what priority they get and what they cost. Any HPC user must study these carefully as they can vary considerably from system to system.

**Post-processing**

In the post-processing part, we have saved the results in 3 formats, Excel files, MATLAB data files, and plots, as presented Fig. 4. The Plots folder contains typical output figures which can be personalised by users, such as electrolyte concentration at the start or end cycles, electrode stoichiometry *etc*. We can also move some key figures from each folder of several cases (like *Sweep_Age_Case_1_4* shown in Fig. 4) to another folder so that we can easily compare one specific figure between different cases. The Excel files correspond to a summary of the results, including all previous inputs and custom outputs, such as loss of capacity, loss of lithium inventory (LLI) to SEI, *etc*. An example of an Excel file for one submitted job (with 4 cases) can be found in *Example of an Excel file for summary* in Supplementary Information. We can also collect the output summary of each job to make a comprehensive summary of all the cases we submitted, which allow us to clearly understand how one specific output is affected by multiple inputs (Fig. 5), further carry out sensitivity analysis, and find the best fit within this round of parameter sweeping.

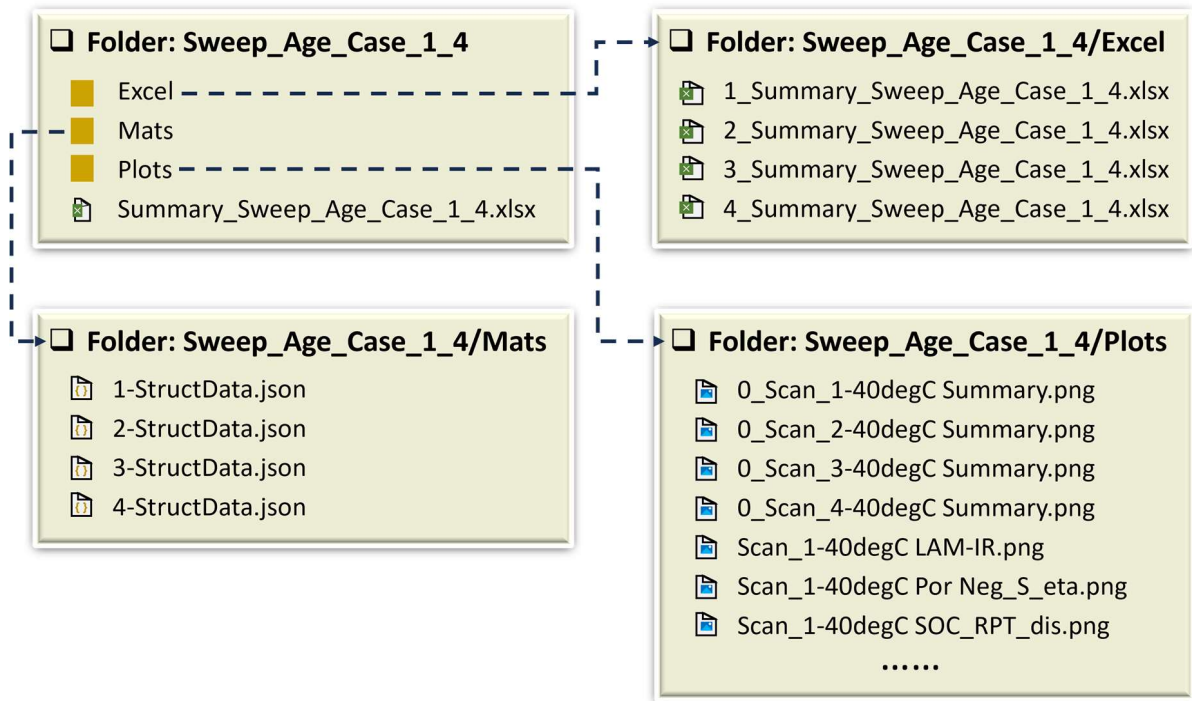

Fig. 4 Example of the output folder.

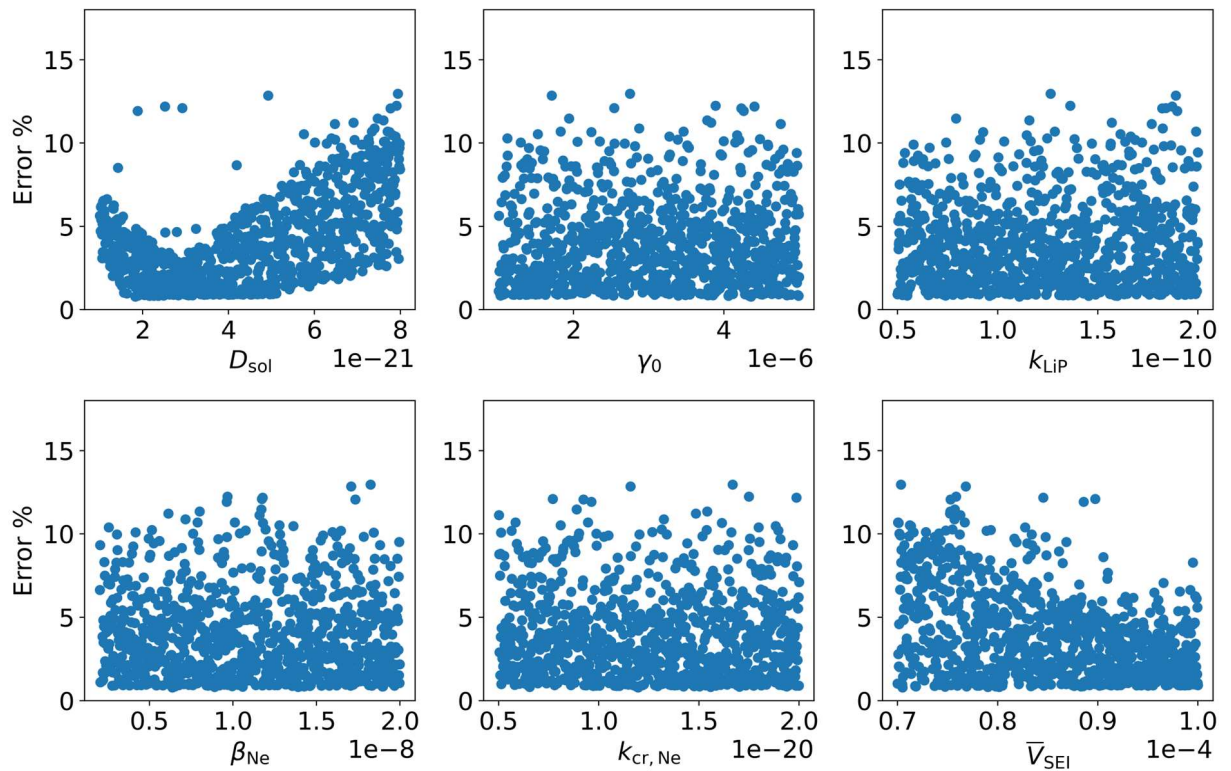

Fig. 5 Mean percentage error between modelling and experimental results as a function of input parameters.

The sensitivity analysis can be conducted if we have a summary of the target input and output relationship, which we have got from the summary Excel files (Summary_Sweep_Age.xlsx in Fig. 3). Fig. 6 shows the total sensitivity indices based on the Sabol's method [32, 33] (see Sensitivity analysis based on Sabol's method for more information), which indicate the decay rate constant of lithium plating $\gamma_0$ is the most sensitive index for the error of the model. Based other value ranking of errors, we can easily pick up the best fit during this round of parameter sweeping, with an serror of 0.79%.

The best fit case in this round (Fig. 7) matches the experimental result up to 3kAh of charge throughput, but then there is some change to the degradation mechanisms occurring inside the LIB which the model fails to capture. Refining this fit is beyond the scope of this work. However, Fig. 7 (b) does offer insight as to which degradation mechanisms dominate for this cell and this set of cycling conditions. SEI growth and electrolyte dry-out caused by SEI growth are responsible for all the capacity fade depicted in Fig. 7 (a), with little lithium plating and negligible mechanical degradation. If a similar cell was designed in the future with excess electrolyte in the cell packaging [6], there would be less dry-out and the cell would last longer. However, other cells may have enough electrolyte already. The power of ParaSweeper is that it can be used to parameterize a degradation model for a specific cell and therefore deliver tailored insights for that cell.

Comparing Figs. 6 and 7 clears up a common misconception around sensitivity analysis. Fig. 6 shows that the model is most sensitive to the parameters concerning lithium plating, but Fig. 7 (b) shows that the best fit to experiment has a very small amount of plating (0.8%). The mechanisms that a degradation model is most sensitive to are not always the mechanisms that are actually occurring inside the LIB. Instead, what sensitivity analysis reveals are the parameters that would make the most difference *if* the relevant mechanisms did occur. In other words, if lithium plating was significant, the simulated SOH vs. charge throughput profile would look very different to the one in Fig. 7 (a) and would not match the experiment. Therefore, the model is highly sensitive to the lithium plating parameters even if those parameters are negligibly small.

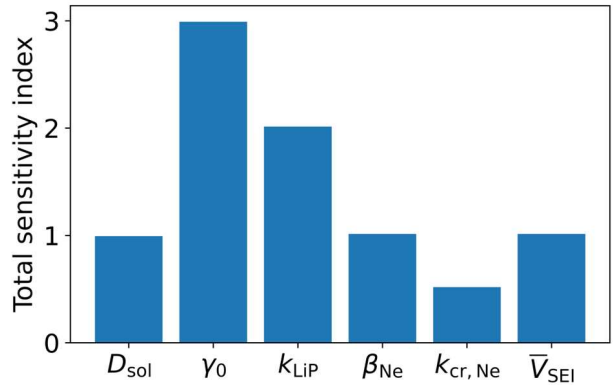

Fig. 6 Sensitivity analysis of mean percentage error with different input parameters.

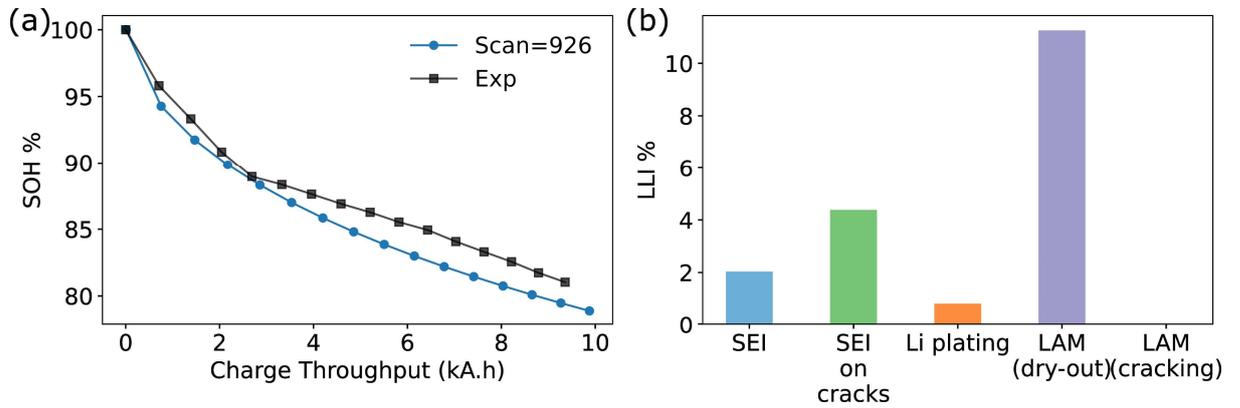

Fig. 7 (a) SOH and (b) loss of lithium inventory (LLI) break down in the best fit of the example sweep.

## Conclusion

In this work, ParaSweeper, a Python script designed to pipeline HTC of physics-based degradation models of LIBs, is introduced. This script incorporates the abundant experience of the authors in debugging physics-based degradation models and manages various pitfalls such as events, cut-offs, solver errors or the solver being stuck in an infinite loop. It also provides guidance on accelerating the simulation and saving computing resources. Based on that, ParaSweeper pipelines the process of getting inputs, running multiple cases simultaneously, and post-processing for later analysis. With these notable features, we showed as an example that by leveraging the powerful high-performance computing platform of Imperial College London, this script can run 1000 cases of virtual ageing experiments within about 17 hours, each case contains 1170 charge/discharge cycles. This corresponds to about one million charge/discharge cycles in total. We have made ParaSweeper public on Zenodo [34] and plan to merge it into the public PyBaMM for the battery modelling community. This script can be a potential game-changer in battery research as it unlocks the power of both HTC and high-fidelity degradation models to enable quick model validation, sensitivity analysis, and reliable diagnosis and prognosis.

# Conflicts of interest

The authors have no conflicts of interest to report.

# Author statement

**Ruihe Li**: Conceptualization, Software, Formal analysis, Visualization, Writing – Original Draft.

**Simon O'Kane**: Validation, Writing – Review & Editing, Supervision.

**Jianbo Huang**: Conceptualization, Software.

**Monica Marinescu**: Investigation, Writing – Review & Editing, Supervision. **Gregory J Offer**: Conceptualization, Writing – Review & Editing, Supervision, Funding acquisition.

# Acknowledgement

The authors would like to acknowledge financial support from EPSRC Faraday Institution Multiscale Modelling project (EP/ S003053/1, grant number FIRG059). The first author is funded as a PhD student by the China Scholarship Council (CSC) Imperial Scholarship.

# Supplementary Information

### Cycling condition

The cycling conditions include reference performance test and ageing test, which repeat in sequence.

*Experiment_RPT   = pb.Experiment(*

  *(# refill*

  *"Hold at 4.2 V until C/100",*

  *"Rest for 1 hours (20 minute period)",*

  *# 0.1C cycle*

  *"Discharge at 0.1C until 2.5 V (30 minute period)",*

  *"Rest for 3 hours (20 minute period)",*

  *"Charge at 0.1C until 4.2 V (30 minute period)",*

"Hold at 4.2V until C/100",

"Rest for 1 hours (20 minute period)",)* 2 )

Experiment_AGE = pb.Experiment((

"Discharge at 1C until 2.5 V",

"Charge at 0.3C until 4.2 V",

"Hold at 4.2 V until C/100",)*78 )

In the above example, the charge and discharge cycle were repeated for 78 times during ageing test. Moreover, certain time periods are assigned in some steps during RPT test, this is an indication for the solver to output the results at these certain points, which can speed up the simulation.

**Example input dictionary**

The following script can be found in Get_sweep.ipynb.

num = 1000; rows_per_file = 4

options_full = {

   'thermal': 'lumped',

   'SEI': 'solvent-diffusion limited',

   'SEI on cracks': 'true',

   'SEI film resistance': 'distributed',

   'SEI porosity change': 'true',

   'particle mechanics': ('swelling and cracking', 'swelling only'),

   'loss of active material': ('stress-driven', 'none'),

   'lithium plating': 'partially reversible'

}

Para_dict = {

   "Total ageing cycles":1170,

   "Ageing cycles between RPT":78,

   "Update cycles for ageing":26,

   "Ageing temperature":[40.0], # 10,25,40

```
        # unchange
        'Initial electrolyte excessive amount ratio':[1.0], # 1.0 or 0.99
        "Cycles within RPT":1,
        "RPT temperature":25,
        "Mesh list":"[5,5,5,60,20]",
        "Para_Set": "OKane2022",
        "Model option":str(options_full),

        ####################### parameter that already there
        # change:
        'Outer SEI solvent diffusivity [m2.s-1]':(1e-21,8e-21),#1e-19~5e-18
        'Dead lithium decay constant [s-1]': (1e-6,5e-6),
        'Lithium plating kinetic rate constant [m.s-1]':(5E-11,2e-10),
        'Negative electrode LAM constant proportional term [s-1]':(2E-9,2E-8),
        'Negative electrode cracking rate':(5e-21,2E-20),
        'Outer SEI partial molar volume [m3.mol-1]':(7e-5,10e-5),
        # unchange:
        "SEI growth activation energy [J.mol-1]":[1e4,],
}
```

**Example of job file used to submit jobs to HPC**

A job file with extension of pbs is created, which contains the following script:

```
#PBS -l walltime=08:00:00
#PBS -l select=1:ncpus=5:mem=80gb
#PBS -J 1-250
module load anaconda3/personal
source activate Env_PBHPC
cd $PBS_O_WORKDIR
```

*python3 Main_sweep_age .py*

In the above file, the first line corresponds to the time needed (8 hours). The second line specifies 5 CPU and 80 GB RAM. The third line specifies the number of the array jobs is 250. The last line tells the script to run is *Main_sweep_age*.py. Within *Main*.py, only 4 cases are submitted. We use the following internal variable in *Main_sweep_age*.py to determine which input file to read and which cases to run.

*i_bundle = int(os.environ["PBS_ARRAY_INDEX"])*

**Example of an Excel file for summary**

Table S1 shows an example of a summary Excel file, which is extracted from Summary_Sweep_Age_Case_1_4.xlsx, some of columns are omitted due to limited space of appearance.) The readers can refer to the attached file in Zenodo [34].

Table S1 Example of summary Excel file

| Scan No | Error % | Dry out | Outer SEI solvent diffusivity [$m^2 \cdot s^{-1}$] | Dead lithium decay constant [$s^{-1}$] | Lithium plating kinetic rate constant [$m \cdot s^{-1}$] | SOH [%] | LLI [%] | LLI to LiP [%] |
|---|---|---|---|---|---|---|---|---|
| 1 | 2.04 | On | 1.44E-21 | 4.32E-06 | 1.36E-10 | 81.29 | 16.15 | 1.18 |
| 2 | 6.49 | On | 1.08E-21 | 1.79E-06 | 8.08E-11 | 87.95 | 10.07 | 0.41 |
| 3 | 9.03 | On | 7.39E-21 | 1.59E-06 | 1.49E-10 | 68.23 | 28.14 | 0.22 |
| 4 | 1.38 | On | 5.44E-21 | 2.63E-06 | 1.42E-10 | 77.58 | 18.85 | 0.43 |
| 5 | 1.3 | On | 4.90E-21 | 1.48E-06 | 1.09E-10 | 77.69 | 18.96 | 0.22 |

**Example notebooks on common problems and methods in the script**

Table S2 Example notebooks on common problems and methods

| Notebook | Purpose |
|---|---|
| 1_cracking event.ipynb | Reproduce the "experiment is infeasible" event due to negative particle crack length larger than particle radius |
| 2_solver_error.ipynb | Reproduce a solver error which can be resolved by increasing *rtol* or make the cycling condition milder |

| 3_charge_at_100SOC.ipynb | Show that if we try to charge the cell at 100%SOC, PyBaMM will skip that step and give an empty solution. In other software, it may return an error or says the experiment is infeasible. |
|---|---|
| 4_Out_of_Range.ipynb | Highlight that many parameters in the physics-based model are measured within a certain range (concentration, T, etc.). Extrapolating out of these ranges may produce unphysical results. |
| 5_Pore_clogging.ipynb | Reproduce pore clogging due to too quick SEI layer growth |
| 6_electrode_sto_out_of_range.ipynb | Reproduce the "experiment infeasible" event due to very low cut-off voltage |
| 7_NonLinear_Ds.ipynb | Reproduce problems brought by non-linear diffusivity |
| 8_save_RAM_time.ipynb | Show how to save memory and time |

Note that the first 7 notebooks are under the scripts/HPC_Li_et_al/CommonProblems folder, while the last one is under folder scripts/HPC_Li_et_al/Methods.

**Sensitivity analysis based on Sabol's method**

Sabol's method on sensitivity analysis was first proposed by I.M Sabol in 2001 [32] to analyse the global sensitivity indices for nonlinear mathematical models. It has been incorporated in an open-source sensitivity analysis package SALib [35]. In this work, the first-order sensitivity index is used. We have adapted the equations summarized in Ref. [33] for the authors' convenience.

Let $\boldsymbol{X} = (X_1, X_2, \ldots X_n)$ and $Y$ be the inputs and output of the model, respectively. Assume that the model output $Y$ can be decomposed into:

$$Y = f(\boldsymbol{X}) = f_0 + \sum_{i=1}^{n} f_i(X_i) + \sum_{i=1}^{n}\sum_{j>i}^{n} f_{i,j}(X_i, X_j) + \cdots + f_{1,2,\ldots n}(X_1, X_2, \ldots X_n), \quad \text{(S-1)}$$

where $f_0$ is a constant, $f_i(X_i)$ are the functions of one input, $f_{i,j}(X_i, X_j)$ are the functions of two inputs, etc. The total variance of the outputs is:

$$V(Y) = \int_0^1 \ldots \int_0^1 f^2(\boldsymbol{X}) d\boldsymbol{X} - f_0^2. \quad \text{(S-2)}$$

The contribution of a generic term $f_{i_1,\ldots i_s}$ ($1 \leq i_1 < \cdots i_s \leq n$) to the total variance is:

$$V_{i_1,\ldots i_s} = \int_0^1 \ldots \int_0^1 f_{i_1,\ldots i_s}^2(X_{i_1}, \ldots X_{i_s}) dX_{i_1}, \ldots X_{i_s}. \quad \text{(S-3)}$$

The ANOVA-like decomposition of the total variance is:

$$V(Y) = \sum_{s=1}^{n} \sum_{i_1 < \cdots i_s}^{n} V_{i_1,\ldots i_s} = \sum_{i=1}^{n} V_i + \sum_{i=1}^{n}\sum_{j>i}^{n} V_{i,j} + \cdots + V_{1,2,\ldots n}. \quad \text{(S-4)}$$

The Sobol's sensitivity indices can be defined as:

$$S_{i_1,\ldots i_s} = \frac{V_{i_1,\ldots i_s}}{V(Y)}, 1 \leq i_1 < \cdots i_s \leq n. \quad \text{(S-5)}$$

The total sensitivity indices of the $i_{\text{th}}$ input can be defined as:

$$S_{T_i} = S_i + S_{i,ci} = 1 - S_{ci}, \tag{S-5}$$

where $S_i$ and $S_{i,ci}$ are the first-order and higher-order index, respectively. $S_{ci}$ is the sum of all the $S_{i_1,\ldots i_s}$ terms that exclude the index $i$.